\documentclass[prl,onecolumn,nofootinbib,aps,tightenlines,preprintnumbers]{revtex4}

\usepackage{epsfig}
\usepackage{amsmath}
\usepackage{hyperref}
\usepackage{fouridx}

  \newcount\hour \newcount\minute
  \hour=\time \divide \hour by 60
  \minute=\time
  \newcommand{\mydate}{\ \today \ - \number\hour :\ifnum \minute<10 0\fi
\number\minute}
\newcommand{\be}{\begin{equation}}
\newcommand{\ee}{\end{equation}}
\newcommand{\bea}{\begin{eqnarray}}
\newcommand{\eea}{\end{eqnarray}}

\newcommand{\gd}{\Gamma_{\rm decay}}
\newcommand{\gbb}{\Gamma_{\rm BB}}
\newcommand{\ew}{{\rm EW}}
\newcommand{\far}{{\rm far}}

\def\Veff{V_\textrm{eff}}
\def\MSbar{$\overline{\textrm{MS}}$ }

\newcommand{\units}[1]{\mathrm{#1}}

\graphicspath{ {../Figures/} }

\begin{document}

\preprint{\hbox{CALT-68-2854}  }

\title{Can the Higgs Boson Save Us From the Menace of the Boltzmann Brains?}

\author{Kimberly K. Boddy and Sean M. Carroll}

\affiliation{California Institute of Technology, Pasadena, CA 91125 }

\begin{abstract}
The standard $\Lambda$CDM model provides an excellent fit to current cosmological observations but suffers from a potentially serious Boltzmann Brain problem.
If the universe enters a de~Sitter vacuum phase that is truly eternal, there will be a finite temperature in empty space and corresponding thermal fluctuations.
Among these fluctuations will be intelligent observers, as well as configurations that reproduce any local region of the current universe to arbitrary precision.
We discuss the possibility that the escape from this unacceptable situation may be found in known physics: vacuum instability induced by the Higgs field.
Avoiding Boltzmann Brains in a measure-independent way requires a decay timescale of order the current
age of the universe, which can be achieved if the top quark pole mass is approximately 178~GeV.
Otherwise we must invoke new physics or a particular cosmological measure
before we can consider $\Lambda$CDM to be an empirical success.
\end{abstract}
\widetext
\maketitle

%%%%%%%%%%%%%%%%%%%%%%%%%%%%%%%%%%%%%%%%%%%%%
\section{Introduction}
Since the discovery of the acceleration of the universe \cite{Riess:1998cb,Perlmutter:1998np}, the $\Lambda$CDM model (cosmological constant, cold dark matter, approximately scale-free primordial perturbations) has provided an excellent fit to a wide variety of data.
However, a problem lurks in the future.
As the universe empties out it approaches a de~Sitter phase, which will exhibit Gibbons-Hawking radiation at a fixed temperature $T_{\rm dS}= H_{*}/2\pi$, where $H_* = \sqrt{\Lambda/3}$ is the constant Hubble parameter \cite{Gibbons:1977mu}.
A three-volume with radius $H_*^{-1}$ can be thought of as a system in thermal equilibrium with entropy $S_* = A_H/4G = 3\pi/G\Lambda$, or $S_* \sim 10^{122}$ for the measured value of $\Lambda$.
Equilibrium systems experience thermal fluctuations that lower the entropy by $\Delta S$ on timescales of order $e^{\Delta S}t_{\rm d}$, and cycle through all allowed states in a Poincar\'e recurrence time $e^{S_*}t_{\rm d}$, where $t_{\rm d}$ is a typical dynamical timescale (the value of which is essentially irrelevant when the entropies under consideration are so large).

This scenario raises the prospect of thermal fluctuations into conditions that seem bizarre from the perspective of conventional thermodynamics \cite{Dyson:2002pf}, which presumes a low-entropy state in the recent past.
These conditions include freak observers or ``Boltzmann Brains''~\cite{Albrecht:2004ke,Bousso:2006xc,Page:2006dt,Linde:2006nw} or, for that matter, any particular local macroscopic configuration of matter.
In such a universe, the overwhelming majority of observers within any specified reference class (including observers in precisely the macrocondition we find ourselves in at the present moment, complete with all the memories and impressions we have of the external world) fluctuate from a higher-entropy past and are surrounded by thermal equilibrium.
This kind of theory is cognitively unstable: it can never simultaneously be true and be justified on the basis of evidence (see {\it e.g.} \cite{albert}).
Such theories are said to have a Boltzmann Brain problem, and $\Lambda$CDM is apparently one of them.

This troublesome situation has nothing to do with speculative ideas about eternal inflation or the cosmological multiverse;
it is a difficulty of \emph{known physics}, or at least the simplest interpretation thereof (a constant vacuum energy, quantum field theory in curved spacetime).
It is therefore worth asking whether there can be any escape from the Boltzmann Brain challenge within known physics.

The simplest solution is if our current vacuum state is unstable and can decay into a different vacuum before Boltzmann Brains form \cite{Page:2006dt}.
Within the Standard Model, sufficiently rapid decay is possible if the Higgs field has another vacuum with a lower energy density.
Interestingly, renormalization-group calculations using current measurements of Standard Model parameters indicate that the Higgs is susceptible to decay to a larger expectation value \cite{Sirlin:1985ux,Sher:1988mj,Casas:1994qy,Espinosa:1995se,Casas:1996aq,Isidori:2001bm,Espinosa:2007qp,ArkaniHamed:2008ym,Ellis:2009tp,EliasMiro:2011aa,Alekhin:2012py,Bezrukov:2012sa,Degrassi:2012ry,Buttazzo:2013uya}.
In this paper we investigate whether this instability is fast enough to prevent the Boltzmann Brain problem from arising.
We find that the answer depends on the precise value of the top quark pole mass and on the cosmological measure--the way in which relative frequencies of events are regularized in an infinite universe (see {\it e.g.} \cite{Winitzki:2006rn,Aguirre:2006ak,Linde:2008xf,SchwartzPerlov:2010ne,Freivogel:2011eg,Salem:2011qz}).
It is apparently necessary to resolve these issues before we can declare $\Lambda$CDM to be a viable theory on its own.

%%%%%%%%%%%%%%%%%%%%%%%%%%%%%%%%%%%%%%%%%%%%%
\section{Measures}
We wish to compare the predicted number of Boltzmann Brains (BBs) to the number of ordinary observers (OOs) in our universe.
We assume that BBs fluctuate into existence in a future de~Sitter phase with some fixed rate $\gbb$ per four-volume.
This rate depends on the details of what kinds of fluctuations are considered, but typical numbers are of the form $\gbb \sim \exp(-10^x)$, where $x$ is between 10 and 100.
For reasonable physical parameters the precise value of $\gbb$ will be irrelevant.
Our universe is plausibly infinite both in space and in time (toward the future), so we need a way of regularizing the numbers of OOs and BBs: a cosmological measure.
Studies of the cosmological measure problem are usually carried out in the context of eternal inflation.
We assume there was no early inflationary phase,
so we consider slight modifications of previous proposals.

The physical situation will depend on the behavior of the potential at large field value $\phi =|\Phi|$, where $\Phi$ is the electroweak Higgs doublet.
The only two vacua are our present one, the electroweak vacuum $\phi_\ew$, and a possible ``far vacuum'' at $\phi_\far$.
How the universe evolves depends on the value $\Lambda_\far$ of the cosmological constant in the far vacuum, where $\Lambda_i = 8\pi GV(\phi_i)$ and $V(\phi_\ew)\approx (2.3\times 10^{-3}\,{\rm eV})^4$.
If $\Lambda_\far > 0$, the far vacuum is also de~Sitter.
In that case there will be thermal fluctuations between the two vacua for all eternity, including up-tunneling \cite{Lee:1987qc,Aguirre:2011ac}.
Over sufficiently large timescales, we expect to see equilibrium statistics, and BBs will dominate.

The interesting cases are therefore when $\Lambda_\far = 0$ or when $\Lambda_\far < 0$.
The first case represents a terminal Minkowski vacuum, in which no further fluctuations occur.
Zero vacuum energy represents an infinite fine-tuning, but we cannot exclude the possibility it being enforced by a symmetry.
The other case is $\Lambda_\far < 0$, by which we include the possibility of a runaway, when there is no vacuum below the Planck scale.
In either case we expect a crunch to a singularity in finite time, so such vacua are also thought of as terminal. We will speak as if spacetime ends at the bubble wall, although it is actually a bit later than that.
(It is possible that a better description of such cases includes a quantum ``bounce'' back to a spacetime description, but we will not consider that possibility here~\cite{Craps:2007ch,Maldacena:2010un}.)

\begin{figure}[t]
  \begin{center}
    \includegraphics[width=0.5\textwidth]{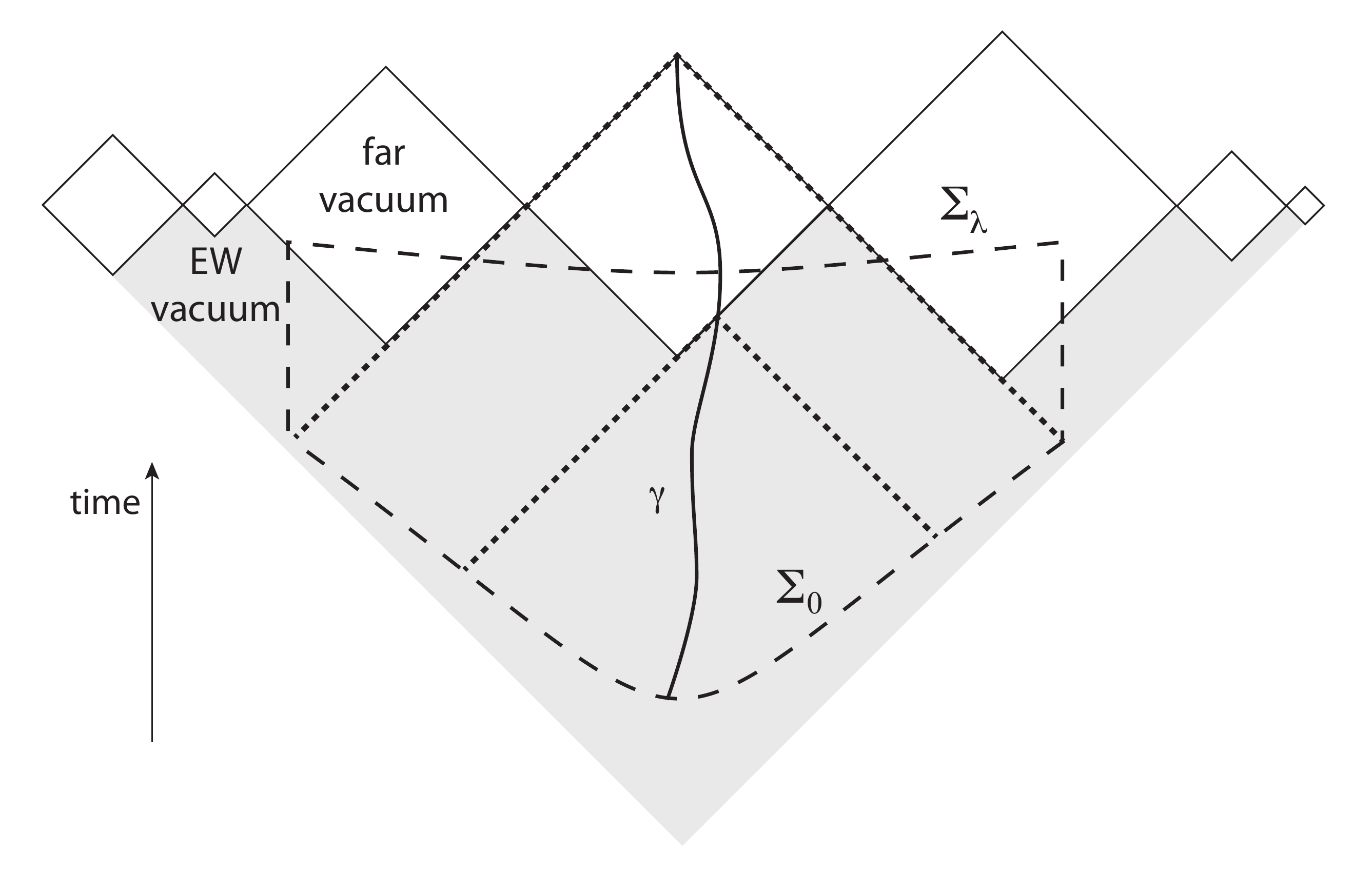}
  \end{center}
  \caption{Conformal spacetime diagram for our universe.
    The shaded region is our electroweak vacuum.
    Diamonds at the top of the diagram are terminal Minkowski vacua; if $\Lambda_\far < 0$, the bottom boundaries of those diamonds represent singularities, and the diamonds themselves are absent.
    Dashed lines depict an initial region $\Sigma_0$ evolving into a later one $\Sigma_\lambda$.
    Dotted lines represent the causal patch of the geodesic $\gamma$, depending on whether $\gamma$ ends at the bubble wall or extends into Minkowski space.}
\label{measure-fig}
\end{figure}

One approach to constructing a measure is to start with some spacelike three-volume $\Sigma_0$ defined at early times, as shown in Figure~\ref{measure-fig}.
We then define a family of hypersurfaces $\Sigma_\lambda$ by extending initially orthogonal geodesics with proper time $\tau$ into the future from $\Sigma_0$.
Each $\Sigma_\lambda$ is the set of all points at some constant parameter $\lambda$, perhaps with some appropriate algorithm to smooth the surfaces, where $\lambda$ is a function of $\tau$.
We calculate the number of OOs and BBs in the four-volume between $\Sigma_0$ and $\Sigma_\lambda$ and take the limit $\lambda\rightarrow\infty$.

Taking $\lambda=\tau$ gives the {\bf proper-time measure}, which naively counts the spacetime volume \cite{Linde:1986fd,Vilenkin:1994ua,Bousso:2007nd}.
If the decay rate $\gd$ to the far vacuum is sufficiently fast that the phase transition percolates,
we spend relatively little time in the electroweak vacuum, and BBs are not produced.
If the transition does not percolate, an infinite amount of volume in the electroweak vacuum is produced; in that case BBs necessarily dominate OOs.
To achieve percolation, the expected lifetime of the electroweak vacuum is of order the actual age of our universe, $\gd \gtrsim H_*^4$ \cite{Page:2006dt}.

In the context of eternal inflation, the proper-time measure has phenomenological problems, such as the
youngness paradox \cite{Tegmark:2004qd,Bousso:2007nd}.
An alternative is the {\bf scale-factor-cutoff measure}, which sets $\lambda$ equal to the scale-factor time along a geodesic congruence, $\lambda = \int H d\tau = \ln{a}$ \cite{Linde:1993nz,Linde:1993xx,Linde:2006nw,DeSimone:2008if}.
In this case the total four-volume living in the electroweak vacuum in the region between the initial hypersurface $\Sigma_0$ and a later hypersurface $\Sigma_\lambda$ is
\be
  U_\ew(\lambda) = \frac{1}{H_*(3 -\kappa)}e^{(3 -\kappa)\lambda}V_0,
\ee
where $V_0$ is the three-volume of $\Sigma_0$ and $\kappa = (4\pi/3)H_*^{-4}\gd$ is the decay probability in a Hubble four-volume.
We see that spacetime volume increases without bound unless $\kappa > 3$, which is just the condition for percolation.
The number of BBs, $N_{\rm BB}(\lambda) = U_\ew(\lambda)\gbb$, therefore also grows without bound,
and will ultimately dominate over OOs.
The scale-factor-cutoff measure, therefore, recovers the same answer as the proper-time measure for our problem:
BBs dominate unless $\gd \gtrsim H_*^4$.

An alternative is to consider the {\bf causal patch measure}~\cite{Bousso:2006ev,Bousso:2007nd}.
The causal patch of a timelike geodesic $\gamma$ extending from $\Sigma_0$ into the future is the intersection of the future of $\Sigma_0$ with the interior of the past light cone of the futuremost point of $\gamma$.
For our problem we can start with any geodesic orthogonal to $\Sigma_0$ and compare the number of OOs to BBs inside the causal patch; no ambiguities due to initial conditions arise.

For the causal patch measure, the result depends on the value of $\Lambda_\far$.
If $\Lambda_\far<0$, the interior of the bubble rapidly crunches to a singularity, ending the geodesic $\gamma$.
In that case we simply want to know whether the geodesic is likely to hit a bubble wall before it observes the formation of a BB; if $\gd > \gbb$, the BB problem is avoided.
If on the other hand $\Lambda_\far=0$, the geodesic enters an asymptotically Minkowski region, and the spacetime volume (in either vacuum) inside the causal patch becomes infinite \cite{Sekino:2010vc}.
In that case BBs will dominate.

To sum up: if the bubble nucleation rate is fast enough that the transition percolates, there is no BB problem.
Otherwise, BBs do dominate according to the proper time or scale factor measures or the causal patch measure with a Minkowski far vacuum.
In the causal patch measure with $\Lambda_\far<0$, BBs are avoided if $\gd > \gbb$.

%%%%%%%%%%%%%%%%%%%%%%%%%%%%%%%%%%%%%%%%%%%%%
\section{Higgs Potential and Decay Rates}
We now investigate the stability and decay rate of the electroweak vacuum, assuming no new physics enters below the Planck scale.
To do so, we need to consider the full structure of the effective potential
\be
  \Veff(\phi) = -\frac{1}{2} m^2 \phi^2 + \frac{1}{4} \lambda \phi^4  + \Delta V (\phi) .
\ee
where $\Delta V$ includes radiative corrections from summing 1PI diagrams with vanishing external momenta~\cite{Coleman:1973jx}.
There is also a $\phi$-independent energy density that should be included, but its effects are negligible at the scales we are interested in for stability~\cite{Casas:1996aq}.
We use the \MSbar renormalization scheme and work in the Landau gauge.
The values of the effective potential at minima are physically meaningful~\cite{Nielsen:1975fs,Fukuda:1975di}, although the potential itself is gauge-dependent \cite{Patel:2011th}.
It is important to verify gauge-independence of the nucleation rate, but we will not delve into this issue here.

The effective potential is known to two loops~\cite{Ford:1992pn,Martin:2001vx}.
The loop approximation is nearly invariant with respect to the renormalization scale $\mu$ if $\mu$ is chosen to minimize the size of radiative corrections~\cite{Casas:1994qy,Casas:1994us}.
Logarithmic terms in $\Delta V$ have the form $\ln(\phi^2/\mu^2)$, so the choice $\mu=\phi$ avoids large logarithms at large $\phi$.
To work consistently at next-to-next-to-leading order (NNLO), the effective potential needs to be improved using the three-loop RG equations with two-loop matching at the electroweak scale.
This procedure correctly resums NNLO logarithm contributions~\cite{Kastening:1991gv,Bando:1992np,Bando:1992wy}.
A recent NNLO calculation of $V_{\rm eff}$ can be found in \cite{Buttazzo:2013uya} (see also \cite{Bezrukov:2012sa,Degrassi:2012ry}).
The most relevant Standard Model parameters are $\alpha_s(M_Z)$ (the strong coupling evaluated at the $Z$ pole mass), the top pole mass $M_t$, and the Higgs pole mass $M_h$.

There are several values to which we would like to compare the electroweak vacuum decay rate $\Gamma_{\rm decay}$.
There is the rate $\Gamma_{\rm fast}$ that is so rapid our vacuum should have already decayed;
the rate $\Gamma_{\rm perc} \sim H_*^4$ below which the phase transition does not percolate;
the rate $\Gamma_{\rm BB}$ at which Boltzmann Brains are created;
and of course $\Gamma_{\rm decay}\rightarrow 0$, where our vacuum becomes stable.
Given the precision of the measured parameters, however, we cannot distinguish between $\Gamma_{\rm fast}$ and $\Gamma_{\rm perc}$, nor between $\Gamma_{\rm BB}$ and $0$.
Similarly, the parameter space for which the far vacuum is de~Sitter but lower energy than our current vacuum is negligibly small.
We are, therefore, interested in two simple questions: is the bubble nucleation rate greater than zero (metastability), and is it fast enough to percolate?

For $\phi \gg \phi_\ew$, we may neglect the $m^2\phi^2$ term and write the effective potential as $\Veff = \frac{1}{4} \lambda_\textrm{eff}(\phi) \phi^4$ \cite{Casas:1994qy,Casas:1996aq}, so that the stability bound is set by requiring $\lambda_\textrm{eff}=0$ at its minimum.
The dividing line between stability and metastability has been calculated as~\cite{Buttazzo:2013uya}
\be
  M_t({\rm GeV}) = 171.4 + 0.5(M_h-125.7) + 357.1(\alpha_s(M_Z)-0.1184) \pm 0.2,
  \label{eq:stabBound}
\ee
where the uncertainty comes from higher order perturbative corrections (and does not include uncertainties in $M_h$ and $\alpha_s$).
If the top pole mass is above this value, the electroweak vacuum decays via bubble nucleation.

Nucleation proceeds via instantons with tree-level Euclidean action $S_0$ and radiative correction $\Delta S$.
Gravitational corrections also become relevant for $R^{-1} \gtrsim 10^{17}~\units{GeV}$ \cite{Isidori:2007vm}.
The action is determined from the bounce solution~\cite{Coleman:1977py} with a characteristic size $R$.
Using the bare potential $\frac{1}{4} \lambda(\phi) \phi^4$ for large $\phi$, we obtain $S_0 = 8\pi^2 /3|\lambda(\mu)|$.
The decay rate per unit volume is then
\begin{equation}
  \Gamma_\textrm{decay} = \frac{1}{R^4} \exp\left[-\frac{8\pi^2}{3|\lambda(\mu)|}
    -\Delta S(\mu R) \right] ,
\end{equation}
where the correction $\Delta S$ has been computed to one loop~\cite{Isidori:2001bm}.
The size of the bounce $R$ will be that which maximizes the decay rate, and $\mu \approx R^{-1}$ is set to minimize the size of the radiative corrections.
The dividing line between percolation and non-percolation has then been calculated as~\cite{ArkaniHamed:2008ym}
\be
   M_t({\rm GeV}) = 178.2 + 0.3(M_h-125.7) + 398(\alpha_s(M_Z)-0.1184) \pm 1.2,
  \label{eq:percBound}
\ee
where the last term is a theoretical error.

\begin{figure}[t]
  \includegraphics[width=0.5\textwidth]{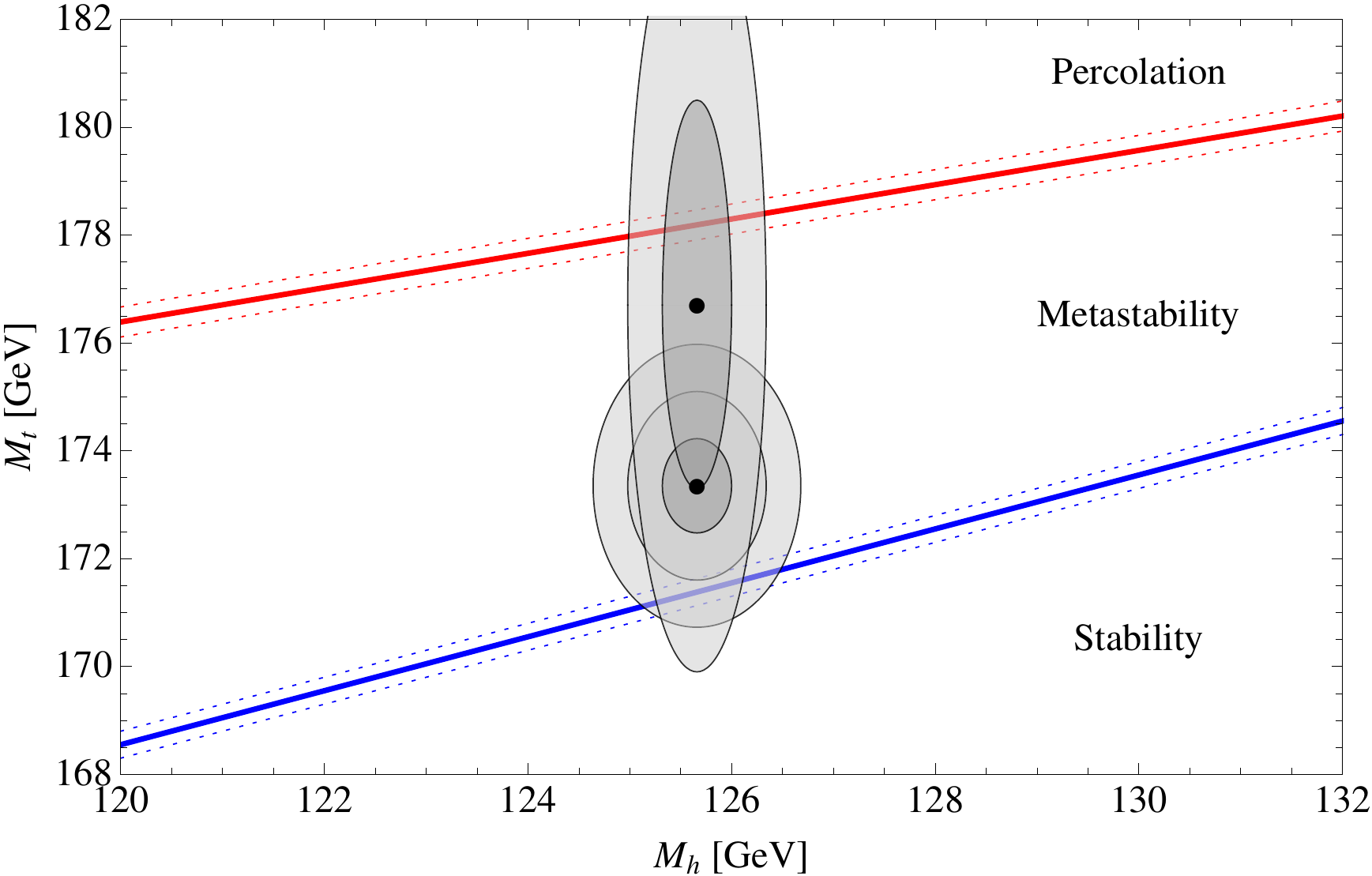}
  \caption{Stability regions for the electroweak vacuum.
    The lower (blue) line is the stability bound in \eqref{eq:stabBound}, and the upper (red) line is the percolation bound in \eqref{eq:percBound}.
    The smaller contours show the $1\sigma$, $2\sigma$, and $3\sigma$ regions, using top mass measurements from the Tevatron \cite{CDF:2013jga}.
    The larger contours show the $1\sigma$ and $2\sigma$ regions, using an alternative determination of the top pole mass from CMS \cite{Chatrchyan:2013haa}.
    The only region that is empirically viable (free of Boltzmann Brains, not already decayed) in a measure-independent way is the upper line, suggesting a top pole mass of $\sim 178$~GeV.}
  \label{fig:stab}
\end{figure}

The bounds from \eqref{eq:stabBound} and \eqref{eq:percBound} are shown in Figure~\ref{fig:stab}, with $\alpha_s(M_Z)=0.1184 \pm 0.0007$ set to its world average value~\cite{Beringer:1900zz}.
The dotted lines near the bounds represent the $1\sigma$ deviation in $\alpha_s$.
Elliptical contours represent measurements on the Higgs mass and top mass.
The Higgs mass $M_h=125.66 \pm 0.34~\units{GeV}$ is obtained from a simple average~\cite{Giardino:2013bma} of measurements from CMS~\cite{CMS-PAS-HIG-13-001,CMS-PAS-HIG-13-002} and ATLAS~\cite{ATLAS-CONF-2013-012,ATLAS-CONF-2013-013}, fitting peaks in the $h\to\gamma\gamma$ and $h\to ZZ \to 4l$ channels.
We use the top mass average of $M_t=173.20 \pm 0.87~\units{GeV}$ from the Tevatron Electroweak Working Group~\cite{CDF:2013jga}, which is consistent with measurements from CMS~\cite{Chatrchyan:2013xza} and ATLAS~\cite{ATLAS-CONF-2013-077}.
These measurements indicate that the electroweak vacuum is metastable, entering into the stability region above the $2\sigma$ level.

There are concerns regarding the methods used to extract the mass of the top~\cite{Alekhin:2012py}.
The top mass parameter that is used in the reconstruction of collider events is not necessarily the pole mass that is needed in the stability calculations.
Furthermore, the top is a confined object that does not exist as an asymptotic state, and non-perturbative effects introduce ambiguities in defining the pole mass.
A way to combat these issues is to extract the \MSbar top mass from the total cross section and use the relation between the \MSbar mass and the pole mass~\cite{Alekhin:2012py,Jegerlehner:2012kn}.
CMS performed this analysis to find a top pole mass of $M_t=\fourIdx{}{}{\; +3.8}{\; -3.4}{176.7}~\units{GeV}$~\cite{Chatrchyan:2013haa}.
While the errors are much larger than the other recent top mass measurements, the quoted value is higher, pushing the electroweak vacuum towards larger decay rates, as shown in Figure~\ref{fig:stab}.

%%%%%%%%%%%%%%%%%%%%%%%%%%%%%%%%%%%%%%%%%%%%%
\section{Conclusions}

$\Lambda$CDM is only viable if Boltzmann Brains are avoided.
There are three possibilities.
One is to invoke an appropriate cosmological measure.
Most measures do not help, but BBs can be avoided in the causal patch measure with a $\Lambda_{\rm far}<0$ terminal vacuum if $\Gamma_\textrm{decay} > \Gamma_\textrm{BB}$.
This criterion amounts to the demand that the potential is in the metastable region, consistent with current measured parameters.

Another possibility is new physics.
Heavy scalar singlets tend to promote stability~\cite{Lebedev:2012zw,EliasMiro:2012ay}, while fermions with large Yukawa couplings promote instability.
For example, a fourth generation with a heavy quark \cite{Sher:1988mj} or right-handed neutrinos with a see-saw mechanism \cite{Casas:1999cd,EliasMiro:2011aa} can destabilize the Higgs potential.
The effects of new particles are model-dependent, and a complete analysis is needed to definitively comment on a particular model.
Alternatively, cosmic acceleration could be due to an ephemeral effect such as quintessence or modified
gravity rather than vacuum energy.

The simplest possibility is if the decay rate of the electroweak vacuum is comparable to the current age of the universe.
We have seen that this scenario can be accomplished without new physics if the top pole mass is $M_t \sim 178~\units{GeV}$, given the measured Higgs mass.
This mass is $5.6\sigma$ larger than the consensus Tevatron/LHC value, although a different determination method might change this result.
It is interesting that the precise value of the top mass plays such an important role in cosmology.

%%%%%%%%%%%%%%%%%%%%%%%%%%%%%%%%%%%%%%%%%%%%%
\section*{Acknowledgments}
We thank Clifford Cheung, Stefan Leichenauer, Michael Ramsey-Musolf, Michael Salem, and Mark Wise for helpful conversations.
This research is funded in part by DOE grant DE-FG02-92ER40701, and by the Gordon and Betty Moore Foundation through Grant 776 to the Caltech Moore Center for Theoretical Cosmology and Physics.

\bibliography{higgs-bb-bib}

\end{document}